\documentclass[12pt,titlepage]{article}
\usepackage{amssymb}
\usepackage[pdftex]{graphicx}
\newcommand{\proof}{{\noindent \bf Proof. }}
\newtheorem{thm}{Theorem}
\newtheorem{defi}{Definition}

\newtheorem{lem}{Lemma}[section]
\newtheorem{cor}[thm]{Corollary}
\newtheorem{prop}[thm]{Proposition}

\newtheorem{rem}{Remark}
\newtheorem{prob}{Problem}

\def\2{\mathbb Z_2}

\newcommand\mbf[1]{\mbox{\boldmath$#1$}}

\title{A generalization of Witsenhausen's zero-error rate for directed graphs
\thanks{This paper is to be presented in part at the
    2014 IEEE International Symposium on Information Theory.}}
\author{{\bf G\'abor Simonyi}\thanks{Research is partially
  supported by the Hungarian Foundation for Scientific Research, Grants K104343
  and K105840.}
$\qquad$\ \  {\bf \'Agnes T\'oth}\thanks{Research is partially supported by the
  Hungarian Foundation for Scientific Research, Grants K104343 and K108947.}\\ 
\medskip \\
Alfr\'ed R\'enyi Institute of Mathematics, \\
Hungarian Academy of Sciences, \\
1364 Budapest, POB 127, Hungary \\ 
\medskip \\
{\tt simonyi.gabor@renyi.mta.hu} \ \ \  {\tt toth.agnes@renyi.mta.hu}}
\date{}
\begin{document}
\maketitle

\begin{abstract}
We investigate a communication setup where a source output is sent through
a free noisy channel first and an additional codeword is sent through a
noiseless but expensive channel later. 
With the help of the second message the decoder should be
able to decide with zero-error whether its decoding of the first message was error-free.
This scenario leads to the definition of a digraph parameter that generalizes
Witsenhausen's zero-error rate for 
directed graphs. We investigate this new parameter for some specific directed
graphs and explore its relations to other digraph parameters like Sperner
capacity and dichromatic number.

When the original problem is modified to require zero-error decoding of the
whole message then we arrive back to the Witsenhausen rate of an
appropriately defined undirected graph.

\end{abstract}

\section{Introduction}

Consider the following situation.
Alice writes a message to Bob consisting of the numbers of several bank
accounts to which Bob has to send some money. She writes in a hurry (she just
got to know that the transfers are urgent if they do not want to pay delay
punishment, but currently she has little time). Therefore her characters are
not very well legible, so Bob may misread some numbers.  However, there are
some rules 
for the possible mistakes, e.g., a $7$ may be thought to be a $1$ but never a
$6$. This relation between the possible digits need not be symmetric: it is
possible that a $0$ is sometimes read as a $6$ but a $6$ may not be decoded
as a $0$. These rules of possible confusions are known both by Alice and by
Bob. 

As Alice is aware of the possibility that Bob misread her message,
later in the day she sends another message to Bob, the goal of which is 
to make Bob certain
whether he read (decoded) the first message correctly or not. If he did
he can transfer the money with complete confidence that he sends it to the
right accounts. If he did not he will know that he does not know the
account numbers correctly and so he better wait and pay the punishment than
transfer the money to the wrong place.

The second message will be received by Bob correctly for certain, 
but it uses an expensive device, e.g., Alice sends it as an sms
from another country after she has arrived there. (Now we understand why she
was in a hurry: she had to arrive to the airport in time.) For some reason, 
every character sent from this foreign country costs a significant amount of
money for her. So she wants to send the shortest possible message that makes
it sure (here we insist on zero-error) that Bob will know whether his decoding
of the original handwritten message was error-free or not. The problem is to
determine the best rate of communication over the second channel as the length
of the original message received tends to infinity.  

\medskip
In Section~\ref{sec:Dil} we describe the abstract communication model for this
scenario and show that the best achievable rate is a parameter of an
appropriate directed graph. We will see that this parameter of a directed graph
is a generalization of the parameter (of an undirected graph) called
Witsenhausen rate. (This means that we also obtain a new interpretation of
Witsenhausen rate.) 

In Section~\ref{sec:bounds} we investigate the relationship with other graph
parameters. These include Sperner capacity and the dichromatic number. 
The former is a generalization of Shannon's graph
capacity \cite{Sha56} to directed graphs. Though originally 
defined to give a general 
framework for some problems in extremal set theory (see \cite{GKV1, GKV2}),
Sperner capacity 
also has its own information theoretic relevance, see \cite{CsN, NR, BLS}. 
The dichromatic number is a generalization of the chromatic number to directed
graphs introduced in \cite{NL}.
Using the above mentioned relations we determine our new parameter for some
specific directed graphs. 

In Section~\ref{sec:comp} we consider a compound channel type version of the
problem parallel to \cite{NR, Witsmulti}.

In Section \ref{sec:extr} some connections to
extremal set theory are pointed out.

In Section \ref{sec:ambit} 
we will consider the setup where the requirement is more ambitious and we
want that Alice's second (the error-free but expensive) message make Bob able
to decode the original message with zero-error. (That is, he will know
the message itself not only the correctness or incorrectness of his original
decoding of Alice's handwriting.) We will see that this
setting leads to the Witsenhausen rate of an undirected graph related to the
problem. This gives a further new interpretation of Witsenhausen rate. 

All logarithms are meant to be of base $2$.

\section{The Dilworth rate of a directed graph} \label{sec:Dil}

\subsection{The communication model} \label{subsec:model}

The abstract setting for our communication scenario is the following. 
We have a source whose output is sent through a noisy channel. (This
belongs to Alice's handwriting.) The input and output alphabets of this channel
are identical and they coincide with the output alphabet of the source.
It is known how the noisy channel can deform the input, in particular we know
what (input) letters can become a certain, possibly different (output) letter
on the other side. (We always assume though that every letter can result in
itself, that is get through the noisy channel without alteration.)  
Later another message is sent (by the same sender) to the same receiver. This
second message is sent via a noiseless channel and 
its goal is to make zero-undetected-error decoding possible, i.e., after
having received this second message the receiver should be able to decide
whether it decoded the first message correctly. The use of the noiseless
channel is expensive, so the second message should be kept as short
as possible.  

Let the shortest possible message that satisfies the criteria have length
$h(t)$ when $t$ characters of the source output are encoded together.
Let $H$ denote the noisy channel. The efficiency of the communication is
measured by the 
quantity $$R_{\rm D}(H):=\liminf_{t\to\infty}\frac{h(t)}{t}$$ 
that we call the Dilworth rate of the noisy channel $H$. (For an explanation of
the name see Remark~\ref{rem:name} in Subsection~\ref{subsect:Dilwdef}.) 
\smallskip

\begin{rem}\label{rem:cvsr}
{\rm Note the special feature of the problem that we characterize a channel by a
rate, that is with a parameter that, unlike channel capacity, we want to be as
small as possible. The reason is that we measure the reliability of a channel
not by the amount of information it can safely transfer but with the amount
of information needed {\em to be added} for making the communication
reliable.} $\Diamond$ 
\end{rem}

\subsection{Dilworth rate and Witsenhausen rate} \label{subsect:Dilwdef}

The relevant properties of $H$ are described by a directed
graph $\vec{G}_H$ having the (common input and output) alphabet as its vertex set and the following
edge set. An 
ordered pair $(a,b)$ of two letters forms a
directed edge of $\vec{G}_H$ if and 
only if $b\neq a$ and the output of $H$ can be $b$ when it is fed by $a$ at
the input. 

\begin{rem}\label{rem:VE}
{\rm As usual we will use $V(\vec{F})$ to denote the vertex set and $E(\vec{F})$ to
denote the edge set of a directed graph $\vec{F}$. We will use similar
notation for undirected graphs that we always consider to be the same as a
{\em symmetrically directed graph}. In such a graph an ordered pair
$(u,v)$ of two vertices forms a directed edge if and only if the
reversely ordered pair $(v,u)$ is also present in the digraph as a directed
edge. We will use the term {\em oriented graph} for directed graphs that do not
contain any edge together with its reversed version. That is $\vec{F}$ is an
oriented graph if $(u,v)\in E(\vec{F})$ implies $(v,u)\notin E(\vec{F})$. As
it is also customary, the term {\em digraph} will be used as a synonym
for ``directed graph''.}
$\Diamond$
\end{rem}
\smallskip

To express $R_{\rm D}$ as a graph parameter we need the following notion.

\begin{defi}
The AND
product $\vec{F}\wedge\vec{G}$ of two directed graphs $\vec{F}$ and $\vec{G}$
is defined as follows.
The vertex set of
$\vec{F}\wedge \vec{G}$ is the 
direct product $V(\vec{F})\times V(\vec{G})$ and vertex $(f,g)$ sends a directed
edge to $(f',g')$ iff either $(f,f')\in E(\vec{F})$ and $(g,g')\in E(\vec{G})$ or $(f,f')\in E(\vec{F})$
and $g=g'$ or $f=f'$ and $(g,g')\in E(\vec{G})$. The $t$-th AND power of a
digraph $\vec{G}$, denoted by $\vec{G}^{\wedge t}$ is the $t$-wise AND
product of digraph $\vec{G}$ with itself. 
\end{defi}

Observe that this graph exponentiation extends to sequences of
letters the 
relation between individual letters $f$ and $f'$ expressing that feeding $f$ 
to the noisy channel $H$ may result in observing letter $f'$ at the
output. 
A sequence of letters at the input of $H$ can result in another such
sequence at the output of $H$ if at each coordinate the character in the
first sequence can result in the corresponding character of the second
sequence. (This includes the possibility that the character does not 
change when sent through $H$.)
\smallskip

\begin{rem} \label{rem:AO}
{\rm The terminology of graph products is not completely standardized. The AND
product we just defined is also called normal product \cite{Berge}, strong
direct product \cite{LLbook}, or strong product \cite{HIK}. We follow the
paper of Alon and Orlitsky \cite{AO} when use the name 
AND product, because we find this name informative. A similar remark applies to
the OR product that we will introduce later in Definition~\ref{defi:OR}.} $\Diamond$
\end{rem}
\smallskip

Recall that the chromatic number $\chi(F)$ of a graph $F$ is the
minimal number of colors that suffice to color the vertices of $F$ so that
adjacent vertices get different colors. If $\vec{F}$ is a digraph, its
chromatic number $\chi(\vec{F})$ is understood to be the chromatic number of
the underlying undirected graph.  
\smallskip

\begin{prop} \label{prop:chi} 
$$R_{\rm D}(H)=\lim_{t\to\infty}\frac{1}{t}\log\chi(\vec{G}_H^{\wedge t}).$$
\end{prop}

\begin{rem}
{\rm It is easy to see that the above limit always exists. (The reason is the
submultiplicative behaviour of the chromatic number under the AND
product).} 
$\Diamond$
\end{rem}

\smallskip
\par\noindent
\proof
Alice and Bob can agree in advance in a proper coloring of $\vec{G}_H^{\wedge
  t}$ with 
$\chi(\vec{G}_H^{\wedge t})$ colors. Alice can send Bob the color of the vertex belonging to
the original source output using $\lceil\log\chi(\vec{G}_H^{\wedge t})\rceil$ bits. Bob
compares this to the color of the vertex representing the sequence he obtained
as a result of his decoding. If the latter color is identical to the one Alice
has sent him, then he can be sure that his decoding was error-free. This is
because any other sequence that could result in his decoded sequence is
adjacent (in $\vec{G}_H^{\wedge t}$) to this decoded sequence, so its color is
different. 

On the other hand, if Alice sent a shorter message through the
noiseless channel, then she could not have $\chi(\vec{G}_H^{\wedge t})$
distinct messages and thus there must exist two adjacent vertices in
$\vec{G}_H^{\wedge t}$ that are encoded to the same codeword by Alice (for the
noiseless channel). Then
one of the two sequences represented by these two adjacent vertices could
result in the other one, while this other one could
also result in itself. Thus Bob cannot make the difference between these two
sequences, one of which is the correct source output sequence while the other
one differs from it. So receiveing this message Bob could not be sure whether
his decoding was error-free or not.
\hfill$\Box$

\medskip
The right hand side expression in Proposition~\ref{prop:chi} can be considered
as a digraph parameter that we will call the Dilworth rate of the digraph
$\vec{G}_H$. 

\begin{defi} \label{defi:Dilrate}
For a directed graph $\vec{G}$ we define its (logarithmic) {\em Dilworth rate}
to be  
$$R_{\rm D}(\vec{G}):=\lim_{t\to\infty}\frac{1}{t}\log\chi(\vec{G}^{\wedge
  t}).$$ The non-logarithmic Dilworth rate is 
$$r_{\rm D}(\vec{G}):=\lim_{t\to\infty}\sqrt[t]{\chi(\vec{G}^{\wedge t}}).$$
Obviously, $R_{\rm D}(\vec{G})=\log r_{\rm D}(\vec{G}).$
\end{defi}
\smallskip

\begin{rem} \label{rem:name}
{\rm Let $\vec{L}$ be the directed graph on $2$ vertices with a single directed
edge. If we consider the vertices of $\vec{L}^{\wedge t}$ as characteristic
vectors of subsets of a $t$-element set then $R_{\rm D}(\vec{L})$ can be
interpreted as the asymptotic exponent of the minimum number of antichains
(sets of pairwise incomparable elements) in
the Boolean lattice of these subsets that can cover all the subsets. The exact
value of this minimum number is given (easily) by a special case of what is
called the ``dual of Dilworth's theorem'' \cite{Dil} (also called Mirsky's
theorem, see \cite{Mir}). This connection to Dilworth's celebrated result is
the reason for calling our new parameter Dilworth rate. Note that the name Sperner
capacity was picked by the authors of \cite{GKV1} for analoguous reasons: the
Sperner capacity of the digraph $\vec{L}$ has a similar relationship with 
Sperner's theorem \cite{Sper}.} $\Diamond$
\end{rem}

\medskip
\par\noindent
The AND product is also defined for undirected graphs. Considering undirected
graphs as symmetrically directed graphs the definition is straightforward. 

Witsenhausen considered the ``zero-error side-information problem'' that led
him to introduce the quantity
$$R_{\rm W}(G)=\lim_{t\to\infty}\frac{1}{t}\log\chi(G^{\wedge t})$$ that is called the
Witsenhausen rate of (the undirected) graph~$G$.  

It is straightforward from the definitions that if $\vec{G}$ is a
symmetrically directed graph and $G$ is the underlying undirected graph (that
we consider equivalent), then
$R_{\rm W}(G)=R_{\rm D}(\vec{G})$. Thus Dilworth rate is indeed a
generalization of Witsenhausen rate to directed graphs. 
\smallskip

\begin{rem} \label{rem:NRfam}
{\rm We note that Nayak and Rose \cite{NR} defines what they call ``the
Witsenhausen rate of a set of directed graphs''. Though formally this gives
the Dilworth rate of a directed graph, the focus of \cite{NR} is elsewhere. 
When its motivating setup results in a family consisting of a
single digraph, then this digraph is symmetrically directed. (See also
Theorem~\ref{thm:wm} in Section~\ref{sec:comp}.)}  
$\Diamond$
\end{rem}

\section{Bounds on the Dilworth rate} \label{sec:bounds}

\subsection{Relation to Sperner capacity and a lower
  bound} \label{subsec:sperner}

Sperner capacity was introduced by Gargano, K\"orner and Vaccaro
\cite{GKV1}. Traditionally this parameter is defined by using the OR product.

\begin{defi} \label{defi:OR}
The OR product $\vec{F}\vee\vec{G}$ of
directed graphs $\vec{F}$ and $\vec{G}$ has vertex set $V(\vec{F})\times V(\vec{G})$ and $(f,g)$ sends
a directed edge to $(f',g')$ iff either $(f,f')\in E(F)$ or $(g,g')\in
E(\vec{G})$. The $t$-th OR power $\vec{G}^{\vee t}$ is the $t$-wise OR product of digraph $\vec{G}$
with itself. 
\end{defi}

Let $\vec{K_n}$ denote the complete directed graph on $n$
vertices, that is the one we obtain from a(n undirected) complete graph $K_n$
when substituting each of its edges $\{a,b\}$ by the two oriented edges
$(a,b)$ and $(b,a)$. The (directed) complement of a digraph $\vec{G}$ is the
directed graph
${\vec{G}}^c$ 
on vertex set $V(\vec{G})$ having edge set
$E({\vec{G}}^c)=E(\vec{K_n})\setminus E(\vec{G}).$

Now we note the
straightforward relation of the AND and OR powers that $({\vec{G}^{\vee
    t}})^c= (\vec{G}^c)^{\wedge t}$.
\medskip

The (logarithmic) Sperner capacity of digraph $\vec{G}$ is defined (see \cite{GKV1, GKV2}) as
$$\Sigma(\vec{G}):=\lim_{t\to\infty}\frac{1}{t}\log\omega_s(\vec{G}^{\vee t}),$$ where
$\omega_s(\vec{F})$ denotes the {\em symmetric clique number}, that is the
cardinality of the largest symmetric clique in 
digraph $\vec{F}$: the size of the largest set $U\subseteq V(\vec{F})$ where for each $f,f'\in
U$ both $(f,f')$ and $(f',f)$ are edges of $\vec{F}$. 
\medskip 

Using the above relation of the AND and OR products, Sperner capacity (of the
complementary graph $\vec{G}^c$) can also be defined as 
$$\Gamma(\vec{G}):=\Sigma({\vec{G}}^c)=\lim_{t\to\infty}\frac{1}{t}\log\alpha(\vec{G}^{\wedge
  t}),$$ 
where $\alpha(\vec{F})$ stands for the independence number (size of the largest
edgeless subset of the vertex set) of graph $\vec{F}$. This is the definition given
in \cite{BLS}. (The authors of \cite{BLS} call this value the Sperner capacity
of $\vec{G}$.) 
\smallskip
\par\noindent
When $G$ is an undirected (or symmetrically directed) graph, then 
$\Gamma(G)=C(G)$, the {\it Shannon capacity} of graph $G$ (see \cite{Sha56}).
\smallskip

We will need a sort of probabilistic refinement of our capacity-like
parameters called their ``within-a-type'' versions, see \cite{CsKtype}. 
First we need the concept of $(P,\varepsilon)$-typical
sequences, cf. \cite{CsK}.
\smallskip

\begin{defi}  \label{defi:peps}
Let $V$ be a finite set, $P$ a probability distribution on $V$, and
$\varepsilon>0$. A sequence ${\mbf x}$ in $V^t$ is said to be $(P,\varepsilon)$-typical if
for every $a\in V$ we have $|{1\over t}N(a|{\mbf x})-P(a)|<\varepsilon$, where
$N(a|{\mbf x})=|\{i: x_i=a\}|$. We denote the set of $(P,\varepsilon)$-typical sequences
in $V^t$ by ${\cal T}^t(P,\varepsilon)$. When $\varepsilon=0$ we also write
${\cal T}_P^t$ for ${\cal T}^t(P,0)$. If ${\mbf x}\in {\cal T}^t(P,0)$ we say that the
{\em type} of ${\mbf x}$ is $P$. 
\end{defi}

\smallskip
\par\noindent
Let $\vec{G}^{\odot t}$ stand for either $\vec{G}^{\wedge t}$ or $\vec{G}^{\vee t}$.
For a directed (or undirected) graph $\vec{F}$ and $U\subseteq V(\vec{F})$ we
denote by $\vec{F}[U]$ 
the digraph induced by $\vec{F}$ on the subset $U$ of the vertex set. We also use the
shorthand notation $\vec{F}^{\odot t}_{P,\varepsilon}=\vec{F}^{\odot t}[{\cal
    T}^t(P,\varepsilon)]$.
\smallskip

Let $\beta(\vec{G})$ be either of the following graph parameters of the
directed graph $\vec{G}$: independence
number, clique number, chromatic number, clique cover number (which is the
chromatic number of the complementary graph), symmetric clique number, or
transitive clique number. (The latter is the size of the largest subset $U$ of
$V(\vec{G})$ 
the elements of which can be linearly ordered so that if $u$ precedes $v$ then
the oriented edge $(u,v)$ is present in $E(\vec{G})$.)

Let the asymptotic parameter $Z(\vec{G})$ be
defined
as $$Z(\vec{G}):=\limsup_{t\to\infty}\frac{1}{t}\log\beta(\vec{G}^{\odot
  t}),$$
while ${Z(\vec{G})}$ stands for $\log z(\vec{G})$. 
  
\smallskip
\par\noindent
\begin{defi} \label{defi:captyp} 
The parameter $Z(\vec{G},P)$ of a digraph $\vec{G}$ within a given type $P$ is the
value $$Z(\vec{G},P)=\lim_{\varepsilon\rightarrow
  0}\limsup_{t\to\infty}\frac{1}{t}\log \beta(\vec{G}^{\odot t}_{P,\varepsilon}).$$  
\end{defi}

We note that for several of the allowed choices of $\beta(\vec{G})$ and $\vec{G}^{\odot t}$
we obtain a graph parameter that already exists in the literature. For example,
when $\beta(\vec{G})=\omega_s(\vec{G})$ and the power we look at is the OR power, we get
Sperner capacity within a given type, that has an important role in the main
results of the papers \cite{GKV1, GKV2}.

If we choose $\beta(\vec{G})=\chi(\vec{G})$ and
the OR power, we obtain the functional called graph entropy, which is defined 
in \cite{KJPrague} and has several nice properties, see \cite{AMSsurv,
  Psurv}, as well as important applications, see e.g. \cite{KK}. When $\beta(\vec{G})=\chi(\vec{G})$ but the exponentiation is the AND power,
then we arrive to the within a type version of Dilworth rate $R_{\rm D}(\vec{G},P)$.
The special case of this for an undirected graph $G$ was already known under
name ``complementary graph entropy'' that could justifiably be
called ``Witsenhausen rate within a given type''. This parameter was
introduced by K\"orner and Longo \cite{KL} 
and further investigated by Marton
\cite{MK}. Although this within-a-type version of Witsenhausen's invariant was
introduced earlier than the non-probabilistic version 
(cf. \cite{KL, Wits}), for the sake of consistancy we denote it by $R_W(G,P)$.

Note that Marton \cite{MK} proved the important
identity $$R_W(G,P)+C(G^c,P)=H(P),$$ where $H(P)=-\Sigma_{i=1}^n p_i\log p_i$
is the entropy of the probability distribution $P=(p_1,\dots,p_n)$. This holds
for any probability distribution $P$ on $V(G)$.
Along the same lines one can also prove the following theorem. We give its
proof for the sake of completeness. 

We will use the notion of fractional
chromatic number $\chi_f(G)$ in the proof. Let $S(G)$ denote the set of
independent sets in $G$. A function $g:S(G)\to R_{+,0}$ is a fractional
coloring of $G$ if for every vertex $v\in V(G)$ we have $\sum_{v\in A\in S(G)} g(A)\ge 1$, that is the sum of the weights $g$ puts on independent sets
containing $v$ 
is at least $1$. (A proper colorig is also a fractional coloring: the color
classes get weight $1$, the other independent sets get weight $0$.) The
fractional chromatic number is $\chi_f(G)=\min_g\sum_{A\in
  S(G)} g(A)$, that is the minimum (taken over all fractional colorings)
of the total weight put on independent sets by a fractional coloring $g$.
(Formally we should write infimum but it is known that the minimum is always
attained. See the book \cite{SchU} for a detailed account on fractional graph
parameters.)

We will need the following properties of the fractional chromatic
number. 
\smallskip

\begin{defi} \label{defi:vtrans}
A directed graph $\vec{G}$ is vertex-transitive if for any two vertices
$u,v\in V(\vec{G})$ it admits an automorphism that maps $u$ to $v$.
\end{defi}
\smallskip

If $F$ is a vertex-transitive graph, then
$\chi_f(F)=\frac{|V(F)|}{\alpha(F)}.$ (For a proof see \cite{SchU},
Proposition~3.1.1 on page 41.)
\smallskip

For every graph $F$ we have
$$\lim_{t\to\infty}\sqrt[t]{\chi(F^{\odot
    t}})=\lim_{t\to\infty}\sqrt[t]{\chi_f(F^{\odot t}}).$$
\par\noindent
The latter follows from Lov\'asz's result \cite{LLfrac} stating that
$\chi(F)\le \chi_f(F)(1+\ln\alpha(F))$ and the obvious inequality
$\chi_f(F)\le\chi(F)$ that holds for all (finite simple) graphs.  

\medskip
\par\noindent
\begin{thm} \label{thm:sumhp}
Let $\vec{G}$ be a directed graph and $P$ an arbitrary fixed probability distribution on
$V(\vec{G})$. Then 
$$R_{\rm D}(\vec{G},P)+\Gamma(\vec{G},P)=H(P).$$
\end{thm}

\proof
Note that by the well-known (and more or less trivial) inequality
$\chi(F)\ge\frac{|V(F)|}{\alpha(F)}$ for every graph $F$, we have 
$\chi(F^{\wedge t}_{P,\varepsilon})\ge\frac{|V(F^{\wedge
    t}_{P,\varepsilon})|}{\alpha(F^{\wedge t}_{P,\varepsilon})}$. Clearly,
this relation also holds if we have a directed graph $\vec{F}$ in place of the
undirected graph $F$. This is straightforward since $\chi(\vec{F})$ and
$\alpha(\vec{F})$ are defined to be identical to the corresponding parameter
of the underlying undirected graph $F$.
It is also
  well-known (cf. e.g. \cite{CsK}) that $\lim_{\varepsilon\to 0}\lim_{t\to\infty}\frac{1}{t}\log
  (|{\cal T}^t(P,\varepsilon)|)=H(P)$. The last two relations immediately give 
$R_{\rm D}(\vec{G},P)+\Gamma(\vec{G},P)\ge H(P).$

\par\noindent
For the reverse inequality let us fix a sequence of probability distributions
$P_t$ on the vertex set of our graph so that $$\lim_{t\to\infty}\max_{a\in
  V(G)}|P(a)-P_t(a)|=0$$ 
\par\noindent
and $$\lim_{t\to\infty}\frac{1}{t}\log\chi(\vec{G}^{\wedge t},P_t)=R_{\rm
  D}(\vec{G},P).$$

Notice that $\vec{G}^{\wedge t}(P_t,0)$ is a vertex-transitive graph, since every
sequence forming an element of ${\cal T}^t(P_t,0)$ can be transformed into any
other such sequence by simply permuting the coordinates.

\par\noindent
Thus $$\begin{array}{rll}
R_{\rm D}(\vec{G},P)=&\lim_{t\to\infty}\frac{1}{t}\log\chi(\vec{G}^{\wedge
  t},P_t)\\ 
=&\lim_{t\to\infty}\frac{1}{t}\log\chi_f(\vec{G}^{\wedge
  t},P_t)\\=&\lim_{t\to\infty}\frac{1}{t}\log\frac{|V(\vec{G}^{\wedge
    t},P_t)|}{\alpha(\vec{G}^{\wedge
    t},P_t)}\\=&\lim_{t\to\infty}\frac{1}{t}\log{|V(\vec{G}^{\wedge 
    t},P_t)|}-\\-&\lim_{t\to\infty}\frac{1}{t}\log{\alpha(\vec{G}^{\wedge
    t},P_t)}\\=&H(P)-\Gamma(\vec{G},P). 
\end{array}$$
\hfill$\Box$
\medskip
\par\noindent
Using standard techniques of the method of types, cf. \cite{Cstype, CsK} we
can already state our lower bound on $R_{\rm D}(\vec{G})$. 
We need the fact that fixing the length $t$ the number of distinct types of a
sequence over some fixed alphabet is only a polynomial
funcion of $t$ (cf. the Type Counting Lemma 2.2 in \cite{CsK}), while the
parameters we 
investigate are asymptotic exponents of some graph parameters that grow
exponentially as $t$ tends to infinity. With this in mind  
we can write $$R_{\rm D}(\vec{G})=\sup_P R_{\rm D}(\vec{G},P),\hskip1cm \Gamma(\vec{G})=\sup_P \Gamma(\vec{G},P).$$
\medskip
\par\noindent
\begin{thm} \label{thm:lowerb}
$$R_{\rm D}(\vec{G})\ge \log |V(\vec{G})|-\Gamma(\vec{G}).$$

\end{thm}

\proof
Using the above equalities, we obtain $R_{\rm D}(\vec{G})+\Gamma(\vec{G})=\sup_P R_{\rm D}(\vec{G},P)+\sup_P
\Gamma(\vec{G},P)\ge \sup_P (R_{\rm D}(\vec{G},P)+\Gamma(\vec{G},P))=\sup_P H(P)=\log|V(\vec{G})|.$ This
gives the lower bound in the statement.
\hfill$\Box$
\smallskip
\par\noindent
Note that Sperner capacity is unkown for many graphs, so the lower bound above
usually does not give a known numerical value. Still, there are some examples
of graphs where Sperner capacity is known and is non-trivial. A basic
example is the cyclically oriented triangle, or more generally, any cyclically
oriented cycle.  
\smallskip
\par\noindent
First we formulate a consequence of the above formula. 
\medskip

\begin{cor} \label{cor:vtrans}
If $\vec{G}$ is a vertex-transitive digraph
then $$R_{\rm D}(\vec{G})=\log|V(\vec{G})|-\Gamma(\vec{G}).$$ 
\end{cor}
\smallskip
\par\noindent
\proof
 Let $P_U$ denote the uniform distribution on the vertex set of
$\vec{G}$. If $\vec{G}$ is vertex-transitive then by symmetry
$R_{\rm D}(\vec{G})=R_{\rm D}(\vec{G},P_U)$ and
$\Gamma(\vec{G})=\Gamma(\vec{G},P_U)$. Combining these 
equalities with Theorem~\ref{thm:lowerb} we obtain
$R_{\rm D}(\vec{G})+\Gamma(\vec{G})=H(P_U)=\log|V(\vec{G})|$ and thus the 
statement. 
\hfill$\Box$
\smallskip

Now we will use this Corollary to determine the Dilworth rate of the cyclically
oriented $k$-length cycle $\vec{C}_k$ for every $k$. 
Note that the complement of a cyclically oriented cycle is a cyclically
oriented cycle of the same length together with all diagonals as bidirected
(or equivalently, undirected) edges. For $k=3$ there are no diagonals, so
the cyclic triangle is isomorphic to its complement. 

The Sperner capacity of the cyclic triangle
was determined in \cite{CFGLS}, cf. also \cite{Bl}, and its value is $\log
2$. This result was generalized by Alon \cite{A}, who proved that the Sperner
capacity of a digraph $\vec{G}$ is bounded from above by
$\log\min\{\Delta_+(\vec{G}),\Delta_-(\vec{G})\}+1,$ 
where $\Delta_+(\vec{F})$ and $\Delta_-(\vec{F})$ stand for the maximum
outdegree and 
maximum indegree of $\vec{F}$, respectively. The indegree and outdegree of a vertex $v$ is
the number of edges at $v$ that are oriented towards or outwards $v$,
respectively. (Cf. \cite{KPS} for a further generalization of Alon's
result.) 

On the other hand,
Sperner capacity is bounded from below by (the logarithm of)
the transitive clique number, the number of vertices in a largest transitively
directed complete subgraph, denoted by $\omega_{\rm tr}(\vec{G})$. 
(This is an easy observation which implies that
substituting $\omega_{\rm s}(\vec{G}^{\vee t})$ 
by $\omega_{\rm tr}(\vec{G}^{\vee t})$ in the definition of Sperner capacity
gives the same value, i.e. it gives an alternative definition of Sperner
capacity, see \cite{Kclg, FKcol, SaSi} and also Proposition 4 and the Remark
following it in \cite{GGKS}.) 
Note that a transitively directed complete
subgraph meant here is not necessarily induced. It is allowed that some
reverse edges are also present on the same subset of vertices. 

\medskip
\par\noindent
\begin{cor} \label{cor:Ck} 
The Dilworth rate of the cyclically oriented $k$-cycle is 
$$R_{\rm D}({\vec{C}_k})=\log\frac{k}{k-1}.$$  
\end{cor}
\smallskip
\par\noindent
\proof
Let the directed complement of $\vec{C}_k$ be denoted by $\vec{S}_k$.
Since $\Delta_+(\vec{S_k})=\Delta_-(\vec{S}_k)=k-2$, Alon's above mentioned
result  
implies that the Sperner capacity of $\vec{S}_k$ is at most $\log (k-1)$.

It is easy to see that $\omega_{\rm tr}(\vec{S}_k)=k-1$, so the lower bound
mentioned above is also $\log (k-1)$.
Since the above two bounds coincide, the Sperner capacity of ${\vec{S}_k}$ is
equal to $\log(k-1)$. 
\par\noindent
Using that ${\vec{C}_k}$ is vertex transitive
Corollary~\ref{cor:vtrans} implies the statement. 
\hfill$\Box$

Note that Corollary~\ref{cor:Ck} shows that
the Dilworth rate is a true generalization of Witsenhausen rate since
$\log\frac{k}{k-1}< \log 2\le R_{\rm W}(C_k)$ if $k\ge 3$.
\smallskip

\begin{defi}\label{defi:acyc}
Call a subset of the vertex set of a directed graph $\vec{G}$ acyclic if it
induces an acyclic subgraph. The latter means that there is no oriented cycle
on these vertices. 
The acyclicity number $a(\vec{G})$ of a directed graph $\vec{G}$ 
is the number of vertices in a
largest acyclic subset of $V(\vec{G})$.
\end{defi}
\smallskip

Note that unlike for
a transitive clique we do not allow reverse edges in an acyclic subgraph. 
\smallskip

Let $m\ge 1$ be an odd number. 
The following tournaments (oriented complete graphs) are also generalizations
of the cyclic triangle. Let $V(\vec{T}_m)=\{0,1,\dots,m-1\}$ and $(i,j)$ is an
edge iff $j-i \equiv r \pmod{m}$ for some $1\le r\le\frac{m-1}{2}$. (Figure 1
shows the tournament $\vec{T}_5$.) Note that it holds for every directed graph
that reversing all of its edges does not change the value of either its
Sperner capacity or of its Dilworth rate. This implies that if $\vec{T}$ is a
tournament then we have $\Sigma(\vec{T}^c)=\Sigma(\vec{T})$ and
$\Gamma(\vec{T}^c)=\Gamma(\vec{T})$. By $\Gamma(\vec{T})=\Sigma(\vec{T}^c)$
all the four values are equal. 

\begin{figure}[h!]
\centering
\includegraphics[scale=0.6]{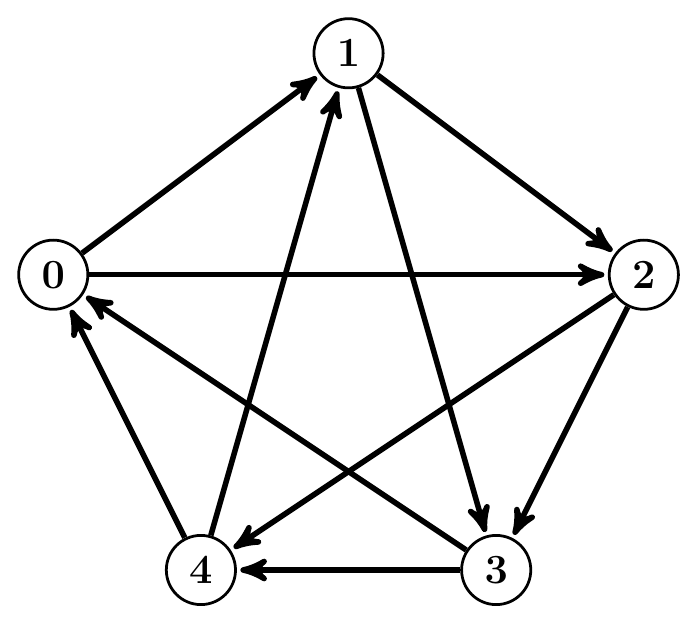}
\caption{The tournament $\vec{T}_5$.}
\label{fig:fc1}
\end{figure}

\begin{cor}\label{cor:tourn}
For all odd integers $m>0$ we have 
$$R_{\rm D}(\vec{T}_m)=\log\frac{2m}{m+1}.$$
\end{cor}

\proof
We know that 
$\Gamma(\vec{G})\ge\log a(\vec{G})$ holds for all directed graphs
(cf. \cite{BLS} and also the discussion before Corollary~\ref{cor:Ck} for an equivalent statement concerning the
complementary digraph). This gives $\Gamma(\vec{T}_m)\ge 
\log\frac{m+1}{2}.$ 

By $\Gamma(\vec{T}_m)=\Sigma(\vec{T}_m)$ (see the note before stating the
Corollary) Alon's result can be applied implying that our lower
bound is sharp. Since $\vec{T}_m$ is 
vertex-transitive we can apply Corollary~\ref{cor:vtrans} to complete the
proof.  
\hfill$\Box$

Observe that Corollary~\ref{cor:tourn} shows not only that the value of the
Dilworth rate of an oriented graph may differ from the Witsenhausen rate of
the underlying 
undirected graph, but also that the difference can be arbitrarily
large. Indeed, denoting the complete graph on $m$ vertices by $K_m$ we have
$\log\frac{2m}{m+1}<\log m=R_{\rm W}(K_m)$ for every $m\ge 2$. The left
hand side of the 
inequality is bounded above by $\log 2$, while the right hand side goes to
infinity with $m$.

\subsection{Dichromatic number and upper bounds} \label{subsec:dichrom}

Now we show that the (logarithm of the) {\em dichromatic number} defined in
\cite{NL} is an upper bound on the Dilworth rate. 
\medskip\par\noindent
\begin{defi} \label{def:dic} 
The {\em dichromatic number} $\chi_{\rm dir}(\vec{G})$ of a directed graph $\vec{G}$ is the
minimum number of acyclic subsets that cover $V(\vec{G})$. A partition of $V(\vec{G})$
into acyclic subsets will be called a {\em directed coloring} or {\em
  dicoloring}. 
\end{defi}
\smallskip

We note that an undirected edge (meaning a bidirected edge) is considered to
be a $2$-length cycle, therefore its two endpoints cannot be both contained
in an acyclic set.
This shows that for undirected (equivalently, symmetrically directed) graphs
the dichromatic number is equal to the chromatic number.  
\smallskip
\begin{rem} {\rm We do not use the term ``acyclic coloring'', because it is
already used for a completely different concept, see \cite{AMcDR}.} $\Diamond$
\end{rem}

\smallskip
\par\noindent
\begin{thm} \label{thm:upb} 
For any directed graph $\vec{G}$ $$r_{\rm D}(\vec{G})\le \chi_{\rm dir}(\vec{G}).$$
\end{thm}
\medskip
\par\noindent
\proof
Let us fix a directed coloring of digraph $\vec{G}$ consisting of $k:=\chi_{\rm dir}(\vec{G})$
acyclic subsets (``color classes''). For each $v\in V(\vec{G})$ let $g(v)$ denote
the color class that contains $v$.

Now consider $\vec{G}^{\wedge t}$. It has $[V(\vec{G})]^t$ vertices. For each sequence
$(a_1,\dots,a_t)\in V(\vec{G}^{\wedge t})$ 
we attach the sequence of colors $(g(a_1),\dots,g(a_t))$. There are $k^t$ such
color sequences, so this gives a partition of $V(\vec{G}^{\wedge t})$ into $k^t$
partition classes. 
\par\noindent
We also give another partition of $V(\vec{G}^{\wedge t})$ according to types. Two
vertices 
are in the same partition class
if their type is the same. By the Type Counting Lemma 2.2 in \cite{CsK}, we
know that this latter partition has at most $(t+1)^{|V(\vec{G})|}$, that is a
polynomial number (in $t$) of classes. Now let ${\cal Q}=(Q_1,\dots,Q_s)$ be the
common refinement of these two partitions. We have $s\le (t+1)^{|V(\vec{G})|}k^t$ by
the foregoing. Now we show that each partition class $Q_i$ induces an
independent set in $\vec{G}^{\wedge t}$. Let two sequences ${\mbf a}=(a_1,\dots,a_t)$
and ${\mbf b}=(b_1,\dots,b_t)$ belong to the same $Q_i$, that is, they have 
the same type and $\forall i: g(a_i)=g(b_i)$. Let $j$ be an index for which
$a_j\ne b_j$. Since $a_j$ and $b_j$ are in the same color class of a valid
dicoloring, we cannot have both  $(a_j,b_j)$ and $(b_j,a_j)$ present in the
graph as a directed edge. If neither is present then there is no edge between 
${\mbf a}$ and ${\mbf b}$. If $(a_j,b_j)\in E(\vec{G})$, then we know that
$(b_j,a_j)\notin E(\vec{G})$, so the oriented edge $({\mbf b},{\mbf a})$ cannot
be an edge of $\vec{G}^{\wedge t}$. We need to show that neither the opposite
oriented edge $({\mbf a},{\mbf b})$ can be present in $\vec{G}^{\wedge t}$. This
is because if $(a_j,b_j)\in E(\vec{G})$, then there should be another
position $\ell$ for which $(b_{\ell},a_{\ell})\in E(\vec{G})$. If this is
true, then  $(a_{\ell},b_{\ell})\notin E(\vec{G})$ and so $({\mbf a},{\mbf
  b})\notin E(\vec{G}^{\wedge t})$. 
The existence of
$\ell$ with the above property holds for the following reason. Consider all
coordinates $h$, for which $g(a_h)=g(a_j)=g(b_j)=g(b_h)$. Denote the set of
these $h$'s $L$. Since vertices $v\in
V(\vec{G})$ with the same ``color'' $g(v)$ induce an acyclic subdigraph, we
can put these vertices into a linear order so, that $(v,v')\in E(\vec{G})$
implies that $v$ precedes $v'$ in this linear order. 
So if $(a_j,b_j)\in E(\vec{G})$, then 
for our edge $(a_j,b_j)$ $a_j$ precedes
$b_j$. However, since ${\mbf a}$ and ${\mbf b}$ has the same type it cannot
happen that for each $h\in L$ $a_h$ precedes $b_h$ in this linear order. So
there must be a coordinate $\ell$ where $b_{\ell}$ precedes $a_{\ell}$ and
this implies the claimed properties. Thus each
partition class $Q_i$ is independent indeed, so the undirected graph
underlying $\vec{G}^{\wedge t}$ can be properly colored with $s\le
(t+1)^{|V(\vec{G})|}k^t$ colors. This implies that 
$r_{\rm D}(\vec{G})=\lim_{t\to\infty}\sqrt[t]{\chi([\vec{G}_H]^{\wedge t})}\le
\liminf_{t\to\infty}\sqrt[t]{(t+1)^{|V(\vec{G})|}k^t}=k=\chi_{\rm dir}(\vec{G}).$
\hfill$\Box$
\medskip
\par\noindent
As a strengthening of the previous theorem we will show that we can also write
a natural fractional relaxation of the dichromatic number on the right hand
side above. (We could not prove this right away, as the
weaker statement will be used in the proof.)
To prove this stronger statement we need some preparation, in
particular we will use the following observations.

\medskip
\par\noindent
First note, that $\chi_{\rm dir}(\vec{F})\le\chi(\vec{F})$ holds for any digraph $\vec{F}$. This is
simply because independent sets in $\vec{F}$ are special acyclic sets, so any proper
coloring of $\vec{F}$ is also a directed coloring of $\vec{F}$.

\medskip
\par\noindent
\begin{prop} \label{prop:submult}
The dichromatic number is submultiplicative with respect to the AND product,
i.e. $$\chi_{\rm dir}(\vec{F}\wedge \vec{G})\le \chi_{\rm dir}(\vec{F})\chi_{\rm dir}(\vec{G}).$$
\par\noindent
In particular, 
$$\chi_{\rm dir}(\vec{F}^{\wedge t})\le [\chi_{\rm dir}(\vec{F})]^t.$$
\end{prop}

\medskip
\par\noindent
A straightforward consequence of Proposition~\ref{prop:submult} is that the
limit $\lim_{t\to\infty}\sqrt[t]{\chi_{\rm dir}(F^{\wedge t})}$ exists.

\medskip
\par\noindent
{\bf Proof of Proposition \ref{prop:submult}} 
Let $c_{\vec{F}}: V(\vec{F})\to \{1,\dots,\chi_{\rm dir}(\vec{F})\}$ and $c_{\vec{G}}: V(\vec{G})\to
\{1,\dots,\chi_{\rm dir}(\vec{G})\}$ be optimal
directed colorings of the digraphs $\vec{F}$ and $\vec{G}$, respectively. 
Using these colorings we define the function $\hat c: V(\vec{F})\times V(\vec{G})\to
\{1,\dots,\chi_{\rm dir}(\vec{F})\chi_{\rm dir}(\vec{G})\}$ as follows. For $(u,v)\in V(\vec{F})\times
V(\vec{G})$ let $\hat c: (u,v)\mapsto (c_{\vec{F}}(u),c_{\vec{G}}(v))$. 
Observe
that the AND product of two acyclic (sub)graphs results in an acyclic
(sub)graph.
Assume for contradiction that $A$ and $B$ are acyclic subsets of $V(\vec{F})$
and $V(\vec{G})$, respectively, and 
$(\vec{F}\wedge \vec{G})[A\times B]$ contains a directed
cycle. (Recall that $\vec{Y}[U]$ denotes the digraph $\vec{Y}$ induces on
$U\subseteq V(\vec{Y})$.)
Let its vertices be $(a_1,b_1),\dots ,(a_k,b_k)$ in the (cyclic) order
the cycle defines, i.e. $((a_i,b_i),(a_{i+1},b_{i+1}))$ is an edge of $\vec{F}\wedge
\vec{G}$ for all $i\in \{1,\dots ,k\}$ where addition is intended modulo $k$. We may
assume without loss of generality that not all $a_i$'s are equal. Then in the sequence $a_1,a_2,\dots,a_k$ we have for all $i\in
\{1,\dots ,k\}$ either $a_i=a_{i+1}$ or $(a_i,a_{i+1})\in E(\vec{F})$ (addition is
again modulo $k$) and for some $i$ the second case occurs. But then there must
be a directed cycle in $\vec{F}[\{a_1,\dots,a_k\}]$ contradicting the assumption
that $A$ is acyclic.
The above implies that $\hat c$ is a directed
coloring of $\vec{F}\wedge \vec{G}$. As it uses $\chi_{\rm dir}(\vec{F})\chi_{\rm dir}(\vec{G})$ colors the
statement is proved.
\hfill$\Box$
\smallskip

\begin{lem} \label{lem:nemno} 
For any digraph $\vec{F}$ and positive integer $k$ we have 
$$r_{\rm D}(\vec{F}^{\wedge k})=[r_{\rm D}(\vec{F})]^k.$$
\end{lem}
\medskip
\par\noindent
\proof 
Fix an arbitrary positive integer $k$. We can write
$$\begin{array}{rll}
r_{\rm D}(\vec{F})=&\lim_{m\to\infty}\sqrt[mk]{\chi(\vec{F}^{\wedge mk})}\\  
=&\lim_{m\to\infty}\sqrt[k]{\sqrt[m]{\chi([\vec{F}^{\wedge k}]^{\wedge m})}}\\
=&\sqrt[k]{\lim_{m\to\infty}\sqrt[m]{\chi([\vec{F}^{\wedge k}]^{\wedge m})}}\\
=&\sqrt[k]{r_{\rm D}(\vec{F}^{\wedge k})},
\end{array}$$ 
\par\noindent
that implies the statement.
\hfill$\Box$

\medskip
\par\noindent
\begin{prop} \label{prop:asympt}
For any digraph $\vec{F}$ we have
$$\lim_{t\to\infty}\sqrt[t]{\chi_{\rm dir}(\vec{F}^{\wedge t})}=r_{\rm D}(\vec{F}).$$
\end{prop}

\par
\noindent
\proof
By $\chi_{\rm dir}(\vec{F})\le \chi(\vec{F})$ we have 
$$\lim_{t\to\infty}\sqrt[t]{\chi_{\rm dir}(\vec{F}^{\wedge t})}\le
\lim_{t\to\infty}\sqrt[t]{\chi(\vec{F}^{\wedge t})}=r_{\rm D}(\vec{F}).$$ 
\par
\noindent
For the reverse inequality we can write  
$$r_{\rm D}(\vec{F})=\lim_{t\to\infty}\sqrt[t]{r_{\rm D}(\vec{F}^{\wedge
    t})}\le\lim_{t\to\infty}\sqrt[t]{\chi_{\rm dir}(\vec{F}^{\wedge t})},$$
\par\noindent
where the equality follows by noticing that Lemma~\ref{lem:nemno} above is
valid for 
all positive integers $t$ and the inequality is a consequence of
$r_{\rm D}(\vec{G})\le\chi_{\rm dir}(\vec{G})$ applied for $\vec{G}=\vec{F}^{\wedge t}$.
\hfill$\Box$

\medskip
\par\noindent
\begin{defi} \label{def:fracdic}
Let the set of subsets of the vertex set inducing an acyclic subgraph in a
digraph $\vec{G}$ be 
${\cal A}(\vec{G})$. A function $g: {\cal A}(\vec{G})\to R_{+,0}$ is called a {\it
  fractional directed coloring} (or fractional dicoloring) if for $\forall v\in V(\vec{G})$ we have
$\Sigma_{v\in U\in{\cal A}(\vec{G})}g(U)\ge 1.$ The {\it fractional dichromatic
  number} of $\vec{G}$ is $$\chi_{{\rm dir},f}(\vec{G})=\min_g\Sigma_{U\in {\cal A}(\vec{G})}g(U),$$
where the minimum is taken over all fractional directed colorings $g$ of $\vec{G}$.
\end{defi}
\smallskip
\par\noindent
Note the obvious inequality $\chi_{{\rm dir},f}(\vec{G})\le\chi_{\rm dir}(\vec{G})$ for any digraph
$\vec{G}$. 
\smallskip

We will need the following lemma.
\smallskip
\par\noindent
\begin{lem} \label{lem:dicfsub}
For any digraphs $\vec{F}$ and $\vec{G}$ we have
$$\chi_{{\rm dir},f}(\vec{F}\wedge \vec{G})\le \chi_{{\rm dir},f}(\vec{F})\chi_{{\rm dir},f}(\vec{G}).$$
\end{lem}
\medskip
\par\noindent
\proof
Let $f$ and $g$ be optimal fractional directed colorings of $\vec{F}$ and $\vec{G}$,
respectively. 
\par\noindent
We use the observation, already verified in the proof of
Proposition~\ref{prop:submult}, stating
that if $A\in {\cal A}(\vec{F})$ and $B\in {\cal A}(\vec{G})$ then the direct
product 
$A\times B$ is in ${\cal A}(\vec{F}\wedge \vec{G})$, i.e. $A\times B$ induces an acyclic
subdigraph in $\vec{F}\wedge \vec{G}$. 

Now give the following weights $w$ to the acyclic sets of $\vec{F}\wedge \vec{G}$. If $H\in
{\cal A}(\vec{F}\wedge \vec{G})$ has a product structure, i.e. $H=A\times B$ for some $A\in
{\cal A}(\vec{F})$ and $B\in{\cal A}(\vec{G})$, then let $w(H)=f(A)g(B)$. If $H$ is not of
this form, then let $w(H)=0$. For any $(a,b)\in V(\vec{F}\wedge \vec{G})$ we have
$\sum_{H\ni (a,b)}w(H)=(\sum_{A\ni a}f(A))(\sum_{B\ni b}f(B))\ge 1,$ thus
$w$ is a fractional dicoloring of $\vec{F}\wedge \vec{G}$. Now we have $\chi_{{\rm dir},f}(\vec{F}\wedge
\vec{G})\le (\sum_{A\in {\cal A}(\vec{F})}f(A))(\sum_{B\in {\cal
    A}(\vec{G})}g(B))=\chi_{{\rm dir},f}(\vec{F})\chi_{{\rm dir},f}(\vec{G}).$ This completes the proof.
\hfill$\Box$
\smallskip

\begin{cor} \label{cor:powfsub}
\par\noindent
For any digraph $\vec{G}$ and any positive integer $t$ we have
$$\chi_{{\rm dir},f}(\vec{G}^{\wedge t})\le [\chi_{{\rm dir},f}(\vec{G})]^t.$$
\hfill$\Box$
\end{cor}

We also need the following result.
\medskip
\par\noindent
\begin{prop} \label{prop:asyfrac}
For any digraph $\vec{F}$ we have
$$\lim_{t\to\infty}\sqrt[t]{\chi_{{\rm dir},f}
(\vec{F}^{\wedge t})}=r_{\rm D}(\vec{F}).$$
\end{prop}

\medskip
\par\noindent
For the proof we need some preparation.

A hypergraph ${\cal H}=(V,{\cal E})$ consists of a vertex set
$V=V({\cal H})$ and an edge set ${\cal E}$, where the elements of ${\cal E}$ are subsets
of $V$. 
A covering of hypergraph ${\cal H}$ is a set of edges the union of which
contains all elements of $V({\cal H})$. 
Let $k({\cal H})$ denote the minimum number of edges in a covering of ${\cal
  H}$.   
A fractional covering of a hypergraph ${\cal H}=(V,{\cal E})$ is a
function $g:{\cal E}\to R_{+,0}$ satisfying for every $v\in V$ that
$\sum_{v\in {E\in {\cal E}}}g(E)\ge 1$. The fractional covering number is
$k_f({\cal H}):=\min_g\sum_{E\in {\cal E}}g(E)$ where the minimization is over
all fractional covers $g$. Clearly, $k_f({\cal H})\le k({\cal H})$. Lov\'asz
proved in \cite{LLfrac} (cf. also \cite{SchU}) that $$k({\cal H})\le k_f({\cal
  H})(1+\log\mu({\cal H})),$$ where $\mu({\cal H})=\max\{|E|: E\in{\cal E}({\cal
  H})\}$, that is the cardinality of a largest edge in ${\cal H}$.

 For a directed graph $\vec{G}$ let ${\cal
  H}_{\vec{G}}=(V(\vec{G}),{\cal E}_{\vec{G}})$ where ${\cal
  E}_{\vec{G}}={\cal A}(\vec{G})$, i.e. it
consists of the acyclic subsets of vertices in $\vec{G}$. It is
straightforward that $k({\cal H}_{\vec{G}})=\chi_{\rm dir}(\vec{G})$ and $k_f({\cal
  H}_{\vec{G}})=\chi_{{\rm dir},f}(\vec{G})$ while $\mu({\cal H})=a(\vec{G})$. 
Thus the above result implies that $$\chi_{\rm dir}(\vec{G})\le\chi_{{\rm
  dir},f}(\vec{G})(1+\log a(\vec{G})).$$

\medskip
\par\noindent
{\bf Proof of Proposition~\ref{prop:asyfrac}}
We have $\lim_{t\to\infty}\sqrt[t]{\chi_{{\rm dir},f}({\vec{F}}^{\wedge t})}\le
\lim_{t\to\infty}\sqrt[t]{\chi_{\rm dir}({\vec{F}}^{\wedge t})}=r_{\rm D}({\vec{F}})$ by Proposition
\ref{prop:asympt} and the obvious inequality $\chi_{{\rm dir},f}(\vec{G})\le \chi_{\rm dir}(\vec{G})$
applied to ${\vec{F}}^{\wedge t}$.
\smallskip
\par\noindent
For the reverse inequality we write
$$\begin{array}{rll}
r_{\rm D}(\vec{F})=&\lim_{t\to\infty}\sqrt[t]{\chi_{\rm dir}(\vec{F}^{\wedge t})}\\
\le&\lim_{t\to\infty}\sqrt[t]{\chi_{{\rm dir},f}(\vec{F}^{\wedge t})(1+\log a(\vec{F}^{\wedge
    t}))}\\
=&\left(\lim_{t\to\infty}\sqrt[t]{\chi_{{\rm dir},f}(\vec{F}^{\wedge t})}\right)\times\\&\times\left(\lim_{t\to\infty}\sqrt[t]
{(1+\log a(\vec{F}^{\wedge t}))}\right)\\
=&\lim_{t\to\infty}\sqrt[t]{\chi_{{\rm dir},f}(\vec{F}^{\wedge t})}.
\end{array}$$
\hfill$\Box$

\begin{thm} \label{thm:frupb}
For any directed graph $\vec{G}$ $$r_{\rm D}(\vec{G})\le \chi_{{\rm dir},f}(\vec{G}).$$
\end{thm}
\medskip
\par\noindent
\proof
From the above we have 
$$\begin{array}{rll}
r_{\rm D}(\vec{G})&=\lim_{t\to\infty}\sqrt[t]{\chi_{{\rm dir},f}(\vec{G}^{\wedge t})}\\&\le
\lim_{t\to\infty}\sqrt[t]{[\chi_{{\rm dir},f}(\vec{G})]^t} =\chi_{{\rm
    dir},f}(\vec{G}).
\end{array}$$
\hfill$\Box$

\par\noindent
There are several directed graphs $\vec{G}$ for which the above upper bound is
sharp. In particular, Corollaries~\ref{cor:Ck}~and~\ref{cor:tourn} can be
proved using Theorem~\ref{thm:frupb} instead of vertex-transitivity. (An
optimal fractional dicoloring of $\vec{T}_5$ is shown on Figure~2.)

\begin{figure}[h!]
\centering
\includegraphics[scale=0.6]{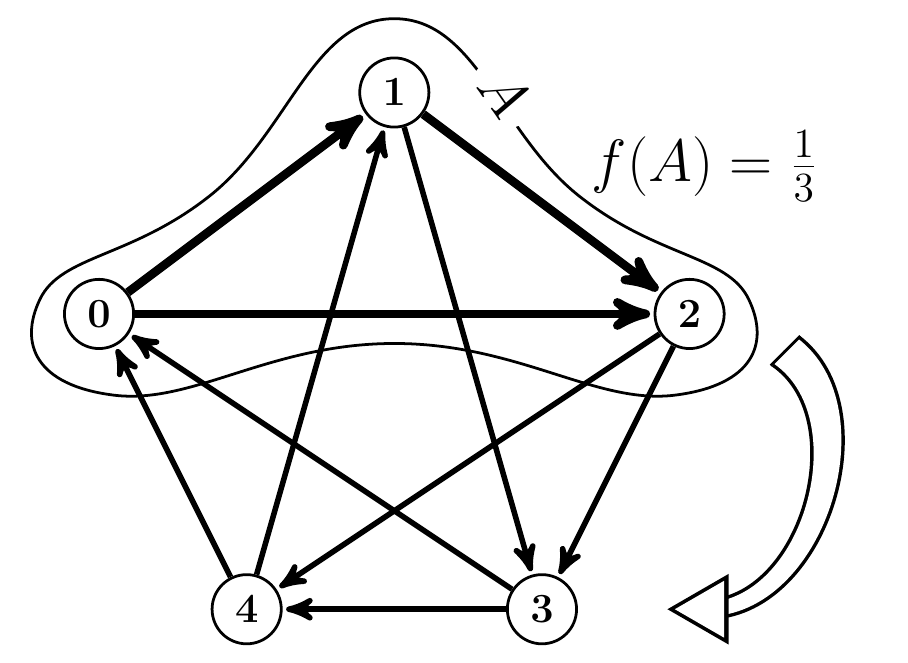}
\caption{An optimal fractional dicoloring of $\vec{T}_5$.}
\label{fig:fc2}
\end{figure}

Note however, that Theorem~\ref{thm:frupb} is not always 
tight: for the graph (symmetrically directed digraph) $C_5$, we have $R_{\rm
  W}(C_5)=\log\sqrt{5}$ by results in \cite{Wits} and 
\cite{LL}, while $\chi_{{\rm dir},f}(C_5)=\frac{5}{2}.$

We present another such example which is not symmetrically directed. 
Let the $5$-length cycle be oriented in an (as much as possible) alternating
manner, that is so that 
only one of its vertices will have outdegree $1$ (implying that two of the $4$
others will have outdegree $2$, and the remaining $2$ have outdegree
$0$). Denote this oriented 
graph by $\vec{A}_5$. (See Figure 3.)

\begin{figure}[h!]
\centering
\includegraphics[scale=0.6]{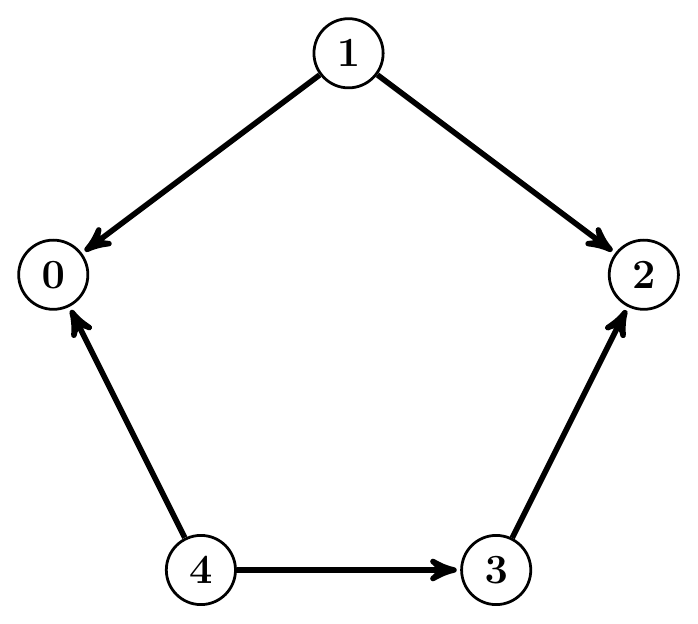}
\caption{The directed graph $\vec{A}_5$.}
\label{fig:fc3}
\end{figure}

We know that $\Sigma(\vec{A}_5)=\log\sqrt{5}$. 
(This is proven as Proposition 4 in \cite{GGKS}, see \cite{SaSi} for more
details on the Sperner capacity of oriented self-complementary graphs. All
other orientations of the $5$-cycle have 
Sperner capacity $\log 2$, see \cite{GGKS} and \cite{KPS}. By
Theorems~\ref{thm:lowerb} and \ref{thm:frupb} this implies that
the Dilworth rate of their complements is $\log\frac{5}{2}$.)   
Thus by Theorem~\ref{thm:lowerb} for the complement of $\vec{A}_5$ we have
$R_{\rm D}(\vec{A}_5^c)\ge\log 5-\log\sqrt{5}$ or equivalently $r_{\rm
  D}(\vec{A}_5^c)\ge\sqrt{5}$. 
(The digraph $\vec{A}_5^c$ is shown on Figure 4.)
\smallskip

\begin{figure}[h!]
\centering
\includegraphics[scale=0.6]{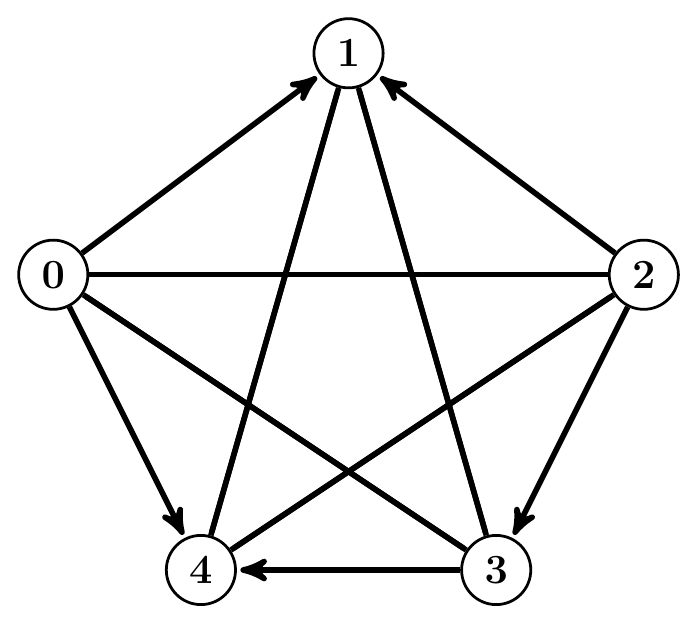}
\caption{The directed graph $\vec{A}_5^c$ (complement of
  $\vec{A}_5$). Bidirected edges are shown as undirected ones.}
\label{fig:fc4}
\end{figure}

\begin{prop}\label{prop:gyok6}
$$\sqrt{5}\le r_D(\vec{A}_5^c)\le\sqrt{6}<\frac{5}{2}=\chi_{{\rm
    dir},f}(\vec{A}_5^c).$$ 
\end{prop}

\proof
The first inequality was already given above. To prove the second inequality
we give $6$ acyclic sets of vertices
in the second power $[\vec{A}_5^c]^{\wedge 2}$ of our graph, that cover all
vertices in $V([\vec{A}_5^c]^{\wedge 2})$. The existence of this covering implies that $r_{\rm
  D}([\vec{A}_5^c]^{\wedge 
  2})\le\chi_{\rm dir}([\vec{A}_5^c]^{\wedge 
  2})\le 6$ thus by Lemma~\ref{lem:nemno} we get $r_{\rm
  D}(\vec{A}_5^c)\le\sqrt{6}$. 

Let us denote the vertices of $V(\vec{A}_5^c)=V(\vec{A}_5)$ by $0,1,2,3,4$ in
their cyclic order, so that in $\vec{A}_5$ we have $d_+(3)=1$ and the
unique outneighbor of $3$ is $2$. (That is, the
outdegree $1$ vertex is $3$, and thus the outdegree $2$
vertices in $\vec{A}_5$ are $4$ and $1$, while $2$ and $0$ have outdegree $0$.)
The following six subsets of $V(\vec{A}_5)\times V(\vec{A}_5)$ induce acyclic 
subgraphs of $V([{\vec{A}_5^c}]^{\wedge 2})$ that entirely cover its vertex set: 

$$44,31,10,23,02;$$
$$14,43,01,30,22$$ $$41,33,32,20$$ $$13,21,42,00$$ $$11,12,04,03$$ $$34,24,40.$$  
One
can check that within all these five sets if $xy$ is to the left of $zw$
in the same line above ($x,y,z,w$ may not all be different),
then either 
$(x,z)$ or $(y,w)$ (or both) form an 
edge of $\vec{A}_5$, thus this is a missing edge in $\vec{A}_5^c$. 
This
implies that as a vertex of $[\vec{A}_5^c]^{\wedge 2}$ the pair $(x,y)$ does not send
an edge to vertex $(z,w)$, therefore the corresponding set of vertices induces
an acyclic subgraph in $[\vec{A}_5^c]^{\wedge 2}$. 
This completes the proof of the second inequality. 

To see that
$\chi_{{\rm dir},f}(\vec{A}_5^c)\ge\frac{5}{2}$ it is enough to realize that any
$3$ vertices of $\vec{A}_5^c$ contains a bidirected edge, thus
any acyclic induced subgraph has at most two vertices. To see equality we can
put weight $\frac{1}{2}$ on all the five $2$-element acyclic subsets.
\hfill$\Box$

\medskip

\begin{rem}\label{rem:coinc}
{\rm Getting the same upper bound for the values determined in
Corrollaries~\ref{cor:Ck}~and~\ref{cor:tourn}
in two different ways above is not pure
coincidence. It follows from the fact that if $\vec{G}$ is 
vertex-transitive then $$\chi_{{\rm
    dir},f}(\vec{G})=\frac{|V(\vec{G})|}{a(\vec{G})}.$$ 
Note that this is a generalization of the relation, that for every
vertex-transitive (undirected) graph $G$  
$$\chi_f(G)=\frac{|V(G)|}{\alpha(G)}$$ that we already referred to right after
Definition~\ref{defi:vtrans} in Subsection~\ref{subsec:sperner}. This latter
equality is presented in \cite{SchU} (Proposition 3.1.1 on page 41) as a
consequence of a more general 
equality concerning vertex-transitive hypergraphs (see Proposition~1.3.4 on page 7
of \cite{SchU}). 

A vertex-transitive hypergraph is a hypergraph that
attains for 
every pair $u,v$ of its vertices an automorphism that maps $u$ to
$v$. Proposition~1.3.4 in \cite{SchU} states that if ${\cal H}=(V,{\cal E})$ is a
vertex-transitive hypergraph then $$k_f({\cal H})=\frac{|V|}{\mu({\cal H})},$$
where (as before; cf. the discussion after stating Proposition~\ref{prop:asyfrac}) $\mu({\cal H})=\max_{E\in{\cal E}}|E|.$
For a directed graph $\vec{G}$ we attach again the hypergraph ${\cal
  H}_{\vec{G}}=(V(\vec{G}),{\cal E}_{\vec{G}})$ where ${\cal
  E}_{\vec{G}}={\cal A}(\vec{G})$. It is
straightforward that if $\vec{G}$ is vertex-transitive then so is  ${\cal
  H}_{\vec{G}}$. The equality quoted for $k_f$ from \cite{SchU} gives the
stated equality $\chi_{{\rm dir},f}(\vec{G})=\frac{|V(\vec{G})|}{a(\vec{G})}$
for vertex-transitive 
digraphs $\vec{G}$.} \hfill$\Diamond$
\end{rem}
\medskip

\section{Compound systems}\label{sec:comp}

Imagine that the handwritten message is left to Bob by one of
his three secretaries but it is not known in advance which one. Their
handwriting is rather different and this has two 
consequences that are important for us. One is that the possible mistakes Bob
can make when decoding the message are different depending on which secretary
wrote him the message. (For example, in the first
secretary's handwriting a $7$ can be thought to be a $1$, while the second
secretary ``crosses'' the leg of $7$, so it can never look like 
a $1$, however it can be confused with a $4$, etc.) This means that in place
of the noisy channel $H$ we had so far, now there are three distinct channels
$H_1, H_2$, and $H_3$ and we do not know in advance which one will be
used. The other important consequence of the secretaries' 
handwriting being different is that Bob recognizes who wrote the message,
i.e., he will know which one of the three noisy channels model the actual
situation. The relevant characteristics (the graphs $G_{H_i}$) of each
of these channels are known by 
Bob and also by his bank. Now it is the bank that will send the second,
error-free but expensive message to Bob. Although the bank knows the
characteristics it does not know which secretary left the first message. So
the second message should make Bob able to decide whether his decoding (of the
first message) 
was correct irrespective of which secretary wrote it. As before, we
are interested (asymptotically) in the shortest possible message the bank can
send to satisfy the requirements. 

Notice that this scenario is basically that of having a
compound channel for the first communication. See \cite{GKV2, NR} for more 
on compound channels from a zero-error point of view.

Here is the abstract setting for the above situation. We have $k$ distinct noisy channels described
by the family ${\cal H}=\{H_1,\dots,H_k\}$. The relevant properties of this
set are
characterised by the family of directed graphs ${\cal\vec{G}}_{\cal
  H}=\{\vec{G}_{H_1},\dots, \vec{G}_{H_k}\}$.  
\smallskip

\begin{defi}\label{defi:comp} {\rm (cf. \cite{NR} and \cite{Witsmulti})}
The Dilworth rate of a family of directed graphs ${\cal\vec{G}}=\{\vec{G}_1,\dots,\vec{G}_k\}$ all
having the same vertex set $V$, is
$$R_{\rm D}({\cal\vec{G}})=\lim_{t\to\infty}\frac{1}{t}\log\chi(\cup_i \vec{G}_i^{\wedge t}),$$ 
where $\cup_i \vec{G}_i^{\wedge t}$ denotes the graph on the common vertex set
$V^t$ 
of the graphs $\vec{G}_i^{\wedge t}$ with edges set $\cup_i
E(\vec{G}_i^{\wedge t})$.
\end{defi}
\smallskip

\begin{prop}\label{prop:multi}
If $m_{\cal H}(t)$ is the shortest possible message the bank should send to
inform Bob about the correctness of his decoding of the handwritten message
for $t$ consecutive rounds, then 
$$\lim_{t\to\infty}\frac{m_{\cal H}(t)}{t}=R_{\rm D}({\cal \vec{G}}_{\cal H}).$$
\end{prop}

\proof
It is enough to prove $m_{\cal H}(t)=\lceil\chi(\cup_i G_{H_i}^{\wedge
  t})\rceil$. 

Let a proper coloring of the graph $\cup_i \vec{G}_{H_i}^{\wedge t}$ be fixed and
agreed on by Bob and the bank in advance. Bob knows that he received the first
message via, say, $H_j$. Since the fixed coloring is a proper coloring of
$\vec{G}_{H_i}^{\wedge t}$, Proposition~\ref{prop:chi} implies that the right hand
side is an upper bound. If $m_{\cal H}(t)$ would be smaller, then there is some
$j$ for which $\vec{G}_{H_j}^{\wedge t}$ has two adjacent vertices for which
the bank sends the same message. If the channel in use is just $H_j$ then
Proposition~\ref{prop:chi} implies that the right hand side above is also a lower
bound. 
\hfill$\Box$
\medskip

The interesting fact about the above quantity is that it is not more then its
obvious lower bound. 
\smallskip

\begin{thm} \label{thm:wm} (\cite{NR}, cf. also \cite{Witsmulti})
For every finite family of directed graphs
${\cal\vec{G}}=\{\vec{G}_1,\dots,\vec{G}_k\}$ we have
$$R_{\rm D}({\cal\vec{G}})=\min_{\vec{G}_i\in\cal{\vec{G}}} R_{\rm
  D}(\vec{G}_i).$$ 
\end{thm}
\smallskip

The analogous result for Witsenhausen rate (that is the special case of the
above when all
graphs are undirected) is proven in \cite{Witsmulti}.  
The above general form is already stated by Nayak and Rose in \cite{NR}
(cf. Remark~\ref{rem:NRfam} of the present paper). They write
that the proof uses essentially the same argument as in \cite{Witsmulti} and
they omit it for the sake of brevity. We do the same. 

\smallskip

\section{Connections to extremal set theory} \label{sec:extr}

As is the case with Sperner capacity, Dilworth rate also has relevance in
extremal set theory. (Recall that both notions got their name from this
relationship, cf. Remark~\ref{rem:name}.) These connections are uncovered when we consider the $t$-length
sequences of vertices of a (di)graph $\vec{G}$ as characteristic vectors of
partitions of a $t$-element set. We already mentioned that if $\vec{G}$ is the
digraph consisting of two vertices and a single oriented edge between them,
then the Dilworth rate is just the asymptotic exponent of the minimum number
of Sperner systems (antichains in the Boolean lattice) that cover all
subsets of a $t$-element set (the elements of the Boolean lattice). This
is known (and easy to prove) to be $t+1$, that is the asymptotic exponent is
$0$. (The situation with Sperner capacity is similar: its value for the above
mentioned single edge graph is the asymptotic exponent of the size of a
largest Sperner system on a $t$-element set which is easy to see to be $1$.) 

\medskip

Here we present another example that we believe to be interesting.     
Let us call a family of pairs of disjoint subsets $(A_i,B_i)$ of a $t$-element
set 
cross-intersecting if for every two pairs $(A_i,B_i)$
and $(A_j,B_j)$ both of  
the intersections $A_i\cap B_j$ and $A_j\cap B_i$ are nonempty. (In other
words, $A_k\cap B_{\ell}=\emptyset$ iff $k=\ell$.) Bollob\'as \cite{Boll}
proved that for such a family $\sum_i\frac{1}{{{|A_i|+|B_i|}\choose |A_i|}}\le
1$. Now we ask, what is the minimum number of cross-intersecting families that
can cover all possible pairs of disjoint subsets of a $t$-element set. If we
are satisfied with determining the asymptotic exponent (i.e. not the exact
value) of this number, then this question is equivalent to asking the
Dilworth rate of an appropriate graph. 
\smallskip

\begin{prop} \label{prop:Bolfam}
Let $B(t)$ denote the minimum number of cross-intersecting families that cover
all pairs of disjoint subsets of a $t$-element set. Then
$$\lim_{t\to\infty}\frac{1}{t}\log B(t)=1.$$
\end{prop}

\proof
Let $\vec{F}$ be the following directed graph. The vertex set of $\vec{F}$ is
$\{0,1,2\}$ and the edge set is $E(\vec{F})=\{(0,1),(1,0),(0,2),(2,0),(1,2)\}$. That
is $\vec{F}$ 
has two undirected (bidirected) edges connecting $0$ to the other two vertices
and one oriented edge from $1$ to $2$. If we encode pairs of disjoint sets of
a $t$-element set by ternary sequences (the positions of $1$'s are the
elements of $A_i$ and the positions of $2$'s are the elements of $B_i$ in
the ternary sequence encoding the pair $(A_i,B_i)$), then it is immediate to
see that $B(t)$ is just the chromatic number of $\vec{F}^{\wedge t}$. Thus 
$R_{\rm D}(\vec{F})$ can indeed be interpreted as the limit in the statement.

Now we have to show that $R_{\rm D}(\vec{F})=1$. We have $\chi_{\rm dir}(\vec{F})=2$, so we
have $R_{\rm D}(\vec{F})\le 1$ by Theorem~\ref{thm:upb}. Since $\vec{F}$ contains an
undirected edge, $\vec{F}^{\wedge t}$ contains a symmetric clique of size
$2^t$. This implies $\chi(\vec{F}^{\wedge t})\ge 2^t$ and thus $R_{\rm D}(\vec{F})\ge
1$. The two inequalities prove $R_{\rm D}(\vec{F})=1.$
\hfill$\Box$

\section{Complete zero-error decoding} \label{sec:ambit}

Here we consider the more ambitious setup, where Bob, otherwise in the same
situation as described in the Introduction, should decode the actual message
with zero-error. (Not only getting to know whether his earlier decoding was correct
or not.) 
\subsection{The closure graph} \label{subsect:clg}

It remains true that all the relevant information to solve this problem is
contained by the directed graph $\vec{G}_H$ defined at the beginning of
Subsection~\ref{subsect:Dilwdef}. We will need the following operation on
directed graphs. 
\smallskip

\begin{defi} \label{defi:hacek}
Let $\vec{F}$ be a directed graph on vertex set $V$. Let the {\em closure
  graph} ${\rm cl}(\vec{F})$ of $\vec{F}$ be the following undirected graph. $$V({\rm cl}(\vec{F})):=V(\vec{F})=V$$
and $$E({\rm cl}(\vec{F})):=\{\{a,b\}: (a,b)\in E(\vec{F})\}\cup$$
$$\cup \{\{a,b\}: \exists v\in V\ {\rm s.t.}\ (a,v),(b,v)\in E(\vec{F})\}.$$
\end{defi}
\smallskip

Note that if $\vec{F}=\vec{G}_H$ then ${\rm cl}(\vec{F})={\rm cl}(\vec{G}_H)$ is the graph
where two vertices $a$ and $b$ are connected if and only if the input letters
they represent can result in the same output letter. This output letter can be
one of $a$ and $b$ but also a third element $v$ of the alphabet. (Recall that
the input and output alphabets of the noisy channel $H$ are identical.) 
The last possibility means that ${\rm cl}(\vec{G}_H)$ may have edges the two
endpoints of which are not adjacent in $\vec{G}_H$ in either direction. 
\medskip

For example, if $\vec{G}_H$ has three vertices, $a,b,c$ and only two
(directed) edges $(a,c)$ and $(b,c)$, then
$\vec{G}_H$ is a bipartite graph, while 
${\rm cl}(\vec{G}_H)$ is the complete (undirected) graph on $3$ vertices. 

\subsection{Relevance of the Witsenhausen rate in this case}

Now we are ready to state the graph theoretic solution of the problem
considered here. 

\begin{thm} \label{thm:W}
Let $h_c(t)$ denote the minimum number of bits Alice should send to Bob
via the noiseless channel for making Bob able to decode a
$t$-length sequence of the source output with zero-error. (The subscript $c$
stands for ``complete''.)
Then $$\lim_{t\to\infty}\frac{h_c(t)}{t}=R_{\rm W}({\rm cl}(\vec{G}_H)),$$ 
the Witsenhausen rate of the closure graph ${\rm cl}(\vec{G}_H)$.  
\end{thm}
\smallskip

\proof
Assume that a $t$-length source output is sent through channel $H$, and the second message sent
by Alice is shorter than $\log\chi([{\rm cl}(\vec{G}_H)]^{\wedge t})$. Then there are
two $t$-length source outputs, that is two sequences ${\mbf x}, {\mbf y}$ in 
$V([{\rm cl}(\vec{G}_H)]^{\wedge t})$ that are adjacent in $[{\rm cl}(\vec{G}_H)]^{\wedge t}$
and for which
Alice sends the same message when encoding either of them for the noiseless
channel. The adjacency of ${\mbf x}$ and ${\mbf y}$ in $[{\rm cl}(\vec{G}_H)]^{\wedge t}$
means that for every $i$ there is a $v_i\in V({\rm cl}(\vec{G}_H))$ such that both
$x_i$ and 
$y_i$ can result in $v_i$ when sent through the noisy channel $H$. (The reason
of this can be that $x_i=y_i=v_i$ or that $(x_i,y_i)$ is an edge of
$\vec{G}_H$, in which 
case $v_i=y_i$ or $(y_i,x_i)$ is an edge of $\vec{G}_H$ and $v_i=x_i$ or 
we have $(x_i,v_i),(y_i,v_i)\in E(\vec{G}_H)$, where $v_i$
differs from both $x_i$ and $y_i$.) Thus if Bob's original decoding of Alice's
(first) message was ${\mbf v}=(v_1,\dots,v_t)$ then he knows that the message
sent could be either of ${\mbf x}$ or ${\mbf y}$. Since Alice's second message
for ${\mbf x}$ is identical to that for ${\mbf y}$, Bob will not know even after
receiving the second message whether the original message was ${\mbf x}$ or
${\mbf y}$.

On the other hand, if the length of Alice's second message is at least
$\log\chi([{\rm cl}(\vec{G}_H)]^{\wedge t})$ then Alice can make Bob able to decide for
sure what the original message was. Indeed, fix a proper coloring of
$[{\rm cl}(\vec{G}_H)]^{\wedge t}$ with $\chi([{\rm cl}(\vec{G}_H)]^{\wedge t})$ colors in advance
that is known by both parties. Encode
each color by a (distinct) sequence of $\lceil\log\chi([{\rm cl}(\vec{G}_H)]^{\wedge
  t})\rceil$ bits. If
the original message was ${\mbf z}=(z_1,\dots,z_t)$ then send Bob the (codeword
for the) color of ${\mbf z}$. Since as a vertex of $[{\rm cl}(\vec{G}_H)]^{\wedge t}$ $
{\mbf z}$ is connected to all those sequences that could result in the same
sequence when sent through $H$ what ${\mbf z}$ can result in, all these
sequences have a different color than ${\mbf z}$ in our coloring of
$[{\rm cl}(\vec{G}_H)]^{\wedge t}$. Thus when Bob gets to know the color of ${\mbf z}$
from Alice's second message he will know that whatever he saw at the output of
$H$ could only arise from ${\mbf z}$ as the input. So he will decode ${\mbf z}$
with zero-error.
\smallskip

Thus we proved that $$h_c(t)=\lceil\log\chi([{\rm cl}(\vec{G}_H)]^{\wedge t})\rceil.$$
So
$\lim_{t\to\infty}\frac{h_c(t)}{t}=\lim_{t\to\infty}\frac{1}{t}\log\chi([{\rm cl}(\vec{G}_H)]^{\wedge t})=R_{\rm W}({\rm cl}(\vec{G}_H))$ as stated.
\hfill$\Box$

\subsection{What graphs can be closure graphs?} \label{subsec:appear}

Not every graph can appear as the closure graph ${\rm cl}(\vec{G})$ of some directed
graph $\vec{G}$. 
\smallskip

\begin{prop}
Let $G$ be a(n undirected) bipartite graph with $|E(G)|\ge |V(G)|+1$. Then $G$
cannot be the closure graph of any directed graph.
\end{prop}

\proof
Let ${\rm cl}(\vec{F})$ be the closure graph of a directed graph $\vec{F}$. 
Observe that if ${\rm cl}(\vec{F})$ has an edge $e$ connecting two vertices that were
not adjacent (in either direction) in $\vec{F}$, then $e$ is contained in a
triangle in ${\rm cl}(\vec{F})$. Let $G$ be a bipartite graph with more edges than
vertices. By bipartiteness $G$ contains no triangle, so if it is a closure
graph of some graph $\vec{G}$, then $\vec{G}$ is just a directed version of
$G$. If any vertex have indegree at least $2$ in $\vec{G}$ that would generate
a triangle in ${\rm cl}(\vec{G})$, so the closure graph
could not be $G$ itself. Since the sum of indegrees equals the number of
edges, we cannot avoid having a vertex with indeegree two if $|E(G)|>
|V(G)|$. This proves the statement. 
\hfill$\Box$

\smallskip

To give a complete characterization of those graphs that can arise
as a closure graph seems
tedious and complicated. It is certainly not a family
of graphs possessing the nice property that it would be closed under taking
induced subgraphs. In fact, the following statement is true.

\begin{prop} \label{prop:allgr}
For any finite simple undirected graph $G$, there exists a directed graph $\vec{F}$
such that ${\rm cl}(\vec{F})$ contains $G$ as an induced subgraph. 
\end{prop}

\proof
Let $G$ be an arbitrary finite simple undirected graph. For every edge
$e=\{a,b\}\in E(G)$ consider a new vertex $v_e$. We add the oriented edges
$(a,v_e)$ and $(b,v_e)$ to our graph $G$. Now delete the edges of $G$ thus
obtaining a graph $\vec{F}$ on vertex set $V(G)\cup \{v_e: e\in E(G)\}$ containing
only the $2|E(G)|$ oriented edges leading to some vertex $v_e$. It is
straightforward to see, that ${\rm cl}(\vec{F})$ contains graph $G$ as an induced
subgraph. 
\hfill$\Box$

\section{
 Open problems}

The general problem concerning the Dilworth rate is to determine it 
for specific directed graphs. Since this is a difficult and mostly open
problem to the related notions of Shannon and Sperner capacities as well as
for the Witsenhausen rate, we cannot expect that this problem is
easy. Nevertheless, we have seen some digraphs for which it was 
solvable (at least when using some non-trivial results already established for
Sperner capacity). Still, there are some directed graphs for which
determining the Dilworth rate seems particularly interesting.
\smallskip

\begin{prob}\label{prob:altot}
What is the Dilworth rate of the graph $\vec{A}_5^c$ we presented in
Subsection~\ref{subsec:dichrom}? Recall that we know $\sqrt{5}\le
r_D(\vec{A}_5^c)\le\sqrt{6}.$
\end{prob}
\medskip

Tournaments play a special role in our setting, because they are exactly those 
oriented graphs the complement of which is also an oriented graph (that is one
without bidirected edges). So it may have some particular interest
how their Dilworth rate behave. 
\smallskip

\begin{prob}
Is there a tournament $\vec{T}$ for which $r_D(\vec{T})$ is strictly smaller
than $\chi_{{\rm dir},f}(\vec{T})$?
\end{prob}
\medskip

\section*{Acknowledgement}
Useful discussions with Imre Csisz\'ar are gratefully acknowledged.

\end{document}